\documentclass[epj]{svjour}
\usepackage{graphics}
\usepackage[title]{appendix}
\usepackage{cite}
\usepackage{amsmath}
\usepackage{breqn}
\usepackage{comment}
\usepackage{graphicx}
\usepackage{caption}
\usepackage{subcaption}
\usepackage[colorlinks=true,citecolor=blue,allcolors=blue]{hyperref}%
\graphicspath{{./graphs/}}

\begin{document}
\title{Inflation using a triplet of Antisymmetric tensor fields}
\author{Abhijith Ajith\inst{1} \and Sukanta Panda\inst{1}
}                     
%
%
\institute{Department of Physics, Indian Institute of Science, Education and Research, Bhopal - 462066, India}
\date{}
%
\abstract{
We study an inflation model driven by a triplet of antisymmetric tensor fields, with minimal and nonminimal couplings to gravity. First, we show that the presence of a triplet of antisymmetric tensor fields can provide inherent background isotropy in the stress-energy tensor contrary to the past studies using an antisymmetric tensor field. Inflation is supported in the presence of non-minimal couplings with gravity. We perform the slow roll analysis and also analyse perturbations to the antisymmetric tensor field as well as the tensor modes of perturbed metric. The speed of gravitational waves manifested from the tensor perturbations is tuned to $c$. We also
study the evolution of the gravitational waves, calculate their power spectrum and tensor spectral index.
%
} 
\maketitle
\section{Introduction}
Inflation is an appealing theoretical paradigm which explores why the universe appears the way it is. Here, it is assumed that an infinitesmally small patch in the early universe underwent rapid exponential expansion leading to the present day observable universe. The theory has been successful in supporting big bang cosmology while posing solutions to the problems such as flatness problem, horizon problem and the monopole problem. Further, inflation gives out a mechanism for the formation of the structures we see in our universe \cite{Riotto:2002yw}. The inflationary expansion excites quantum fields and extends their perturbations from quantum scales to cosmological ranges. Energy density fluctuations are generated as these fluctuations grow into classical scales. Once these classical perturbations reenter the observable universe, they can induce matter and temperature anisotropies. This is believed to be the reason for the generation of the observed anisotropy in CMB and the distribution of large scale structures. Therefore, inflationary model building, has been of great research interest in recent decades. Generally, inflation models either incorporate external driving fields or modifications to gravity \cite{Starobinsky:1980te,Inagaki:2019hmm,Zhang:2021ppy,Sangtawee:2021mhz,Bamba:2015uma,Bhattacharjee:2021kar}. Models incorporating single or multiple scalar fields are quite popular in the scientific literature  \cite{Riotto:2002yw,Starobinsky:1985ibc,Abedi:2016sks,Kodama:2021yrm,Wands:2007bd,Gong:2006zp,Ohashi:2011na,Vazquez:2018qdg,Bartolo:2021wpt}. 

The observational data from the CMB is used to test the authenticity of such models. The recent Planck CMB data  \cite{Iacconi:2019vgc,Planck:2018jri,Planck:2019kim} either rules out or applies tight restrictions to many of the conventional models. In addition, the swampland conjectures from the string theory have put additional constraints on these scalar field models 
\cite{Andriot:2018mav,Brennan:2017rbf,Obied:2018sgi,Garg:2018reu,kinney2019,Kallosh:2019axr}. Despite the rich abundance of scalar field models driven by single and multiple fields, the statistical anomalies in the CMB such as anisotropy and dipolar asymmetry still remain unanswered \cite{Planck:2018jri}. This has encouraged researchers to look for alternative models incorporating vector and tensor fields \cite{Golovnev:2008cf,Darabi:2014aaa,Bertolami:2015wir,Emami:2016ldl,Koh:2009ne,Aashish:2018lhv,Aashish:2019zsy,Aashish:2020mlw,Paul:2020duu,Golovnev:2011yc,Rodriguez:2015rua}.

Vector field models suffer from generic instabilities like ghost and gradient instabilities \cite{Himmetoglu:2009qi,BeltranJimenez:2017cbn,Golovnev:2011yc}. However, recent studies of inflation with multiple vector fields have shown advancements over the previous ones \cite{Gorji:2020vnh,Murata:2021vnb,jiro2009,jiro2010a}.
Speaking of tensor fields, several studies on inflation supported by nonsymmetric tensors, particularly the antisymmetric tensor, exist in the scientific literature \cite{Prokopec:2005fb,Koivisto:2009sd,Obata:2018ilf,Elizalde:2018rmz}. The presence of antisymmetric tensor in the early universe is motivated by the superstring models \cite{Rohm:1986,Ghezelbash:2009gf,jiro2010b}. Inflationary cosmology employing $n-$forms was discussed in Refs. \cite{jiro2010b,jiro2013,jiro2015,2012JCAP...12..016M,DeFelice:2012jt}. 
 n-forms having couplings with gravity are able to mimic slow roll inflation. In 4 dimensions, the 3-form and 0-form have one degree of freedom, where as the 1-form and 2-form have two degrees of freedom. In this sense, 2-form and 3-form models can be said to be duals to vector and scalar theories respectively. The detailed analysis of the background dynamics for the generic n-form inflation model along with the perturbative evolutions for the 0, 1, 2, 3-forms are studied in \cite{Germani:2009iq,Koivisto:2009sd}. At linear level, the inflation model with 2-form fields encounters instabilities similar to vector field model.  For the  3-form field, inflation can be attained without even necessitating slow roll and can be freed from instabilities \cite{Koivisto:2009sd,DeFelice:2012jt}. However, our emphasis is to show the inflationary solutions attained in the presence of an antisymmetric tensor field.

Antisymmetric tensor fields of rank 2 and 3 acting as a standalone driving field for inflation was studied in Refs. \cite{Koivisto:2009ew,Koivisto:2009sd}, where generic instabilities were highlighted. Afterwards, it was shown models with a driving rank-2 antisymmetric tensor field can give inflationary solutions in the presence of minimal and non-minimal couplings with gravity \cite{Aashish:2021gdf,Aashish:2018lhv,Aashish:2019zsy,Aashish:2020mlw}. These models are devoid of the generic ghost and gradient instabilities unlike vector field and 2-form inflation models studied earlier. Further, they predict a nearly scale invariant power spectrum for the tensor perturbations in quasi de-Sitter limit \cite{Aashish:2020mlw,Aashish:2019zsy,Aashish:2021gdf}.

A viable model of cosmic inflation should have stable de-Sitter solutions, and should be able to support  60-70 e-folds in the quasi de-Sitter limit \cite{Dodelson:2003ft}. A suitable choice for the background metric is the flat FLRW metric which obeys the principles of homogeneity and isotropy. For the $(-,+,+,+)$ signature, our metric reads,
\begin{equation}\label{1.1}
     g_{00}=-1, \hspace{.5cm}  g_{ij}=a(t)^{2}\delta_{ij}
 \end{equation}
The past studies on antisymmetric tensor field inflation employed an antisymmetric tensor field $B_{\mu\nu}$ to drive inflation. Following the principles of homogeneity and isotropy, the stress energy tensor obtained from the $B_{\mu\nu}$ lagrangian should have a diagonal structure. But this is not inherent in those models and isotropy is ensured by letting the non-diagonal components go to 0 \cite{Aashish:2018lhv,Aashish:2019zsy,Aashish:2020mlw,Aashish:2021gdf}. This puts additional constraints on the models. In our work, we try to get around this inconvenience by using a triplet of mutually orthogonal antisymmetric tensor fields. The use of three mutually orthogonal fields for achieving isotropy has been employed in vector field models, similar strategy can be deployed in 2-form field model also \cite{Tishue:2022vwc,Gorji:2020vnh,PhysRevD.85.123504,Funakoshi_2013}. Owing to the antisymmetric nature of the fields, each field in general has 6 independent components. We have the freedom to choose the background structure of these fields. As a preliminary study, we are looking at a simple form. Our choice of the background fields reads as,
\begin{equation} \label{1.2}
            B_{0i}^{(k)}=0, \hspace{2em} B_{ij}^{(k)}=B(t){\epsilon_{ijk}}
\end{equation} 
where $\epsilon_{ijk}$ is the completely antisymmetric levicivita tensor of rank 3. Here k runs from 1 to 3 and each k defines a  different antisymmetric field.

The organization of this paper is as follows. In section \ref{sec:2} we setup the minimal model. In Section \ref{sec:3} we develop the non-minimal model, and perform the slow roll analysis in section \ref{sec:4}. In section \ref{sec:5}, we look at the energy conditions in general relativity in the context of our model. The possible perturbations in our theory are discussed in section \ref{sec:6}. We conclude our findings and address the future possibilities in section \ref{sec:7}.

\section{The Minimal model}\label{sec:2}
We begin our analysis employing the simple minimal model without any explicit couplings of the tensor fields with gravity. The action is written as, 
\begin{multline} \label{2.1}
          S=\int d^{4}x\sqrt{-g} \ [\frac{R}{2\kappa}-\frac{1}{12}H^{(k)}_{\lambda\mu\nu}H^{(k)\lambda\mu\nu}+\\ \frac{\tau}{2}(\nabla_{\lambda}B^{(k)\lambda\nu})(\nabla_{\mu}{B^{(k)\mu}}_{\nu})-V(B)]
\end{multline}
Here $g$ is the metric determinant, $R$ is the Ricci scalar, and $\kappa$ is the inverse of squared planck mass. $\tau$ is a dimensionless parameter. The rank three tensor $H^{(k)}_{\lambda\mu\nu}$ is defined for each of the three $B$ fields as, $H^{(k)}_{\lambda\mu\nu}=\nabla_{\lambda}B^{(k)}_{\mu\nu}+\nabla_{\nu}B^{(k)}_{\lambda\mu}+\nabla_{\mu}B^{(k)}_{\nu\lambda}$. In the action, the first term in the parentheses corresponds to Einstein-Hilbert term. The second term constituting the $H$ tensor is the gauge invariant kinetic term corresponding to the transformation
\begin{equation}
B^{(k)}_{\mu\nu}\rightarrow B^{(k)}_{\mu\nu}+\partial_{\mu}\Lambda_{\nu}-\partial_{\nu}\Lambda_{\mu}    
\end{equation}
The presence of the third term in the action is important to remove the instabilities while looking at the perturbations in the field. As evident from Refs. \cite{Aashish:2018lhv,Aashish:2019zsy}, the $\tau$ term in the action gives dynamic nature to all the perturbed modes and gets rid of ghost and gradient instabilities. $V(B)$ is the potential term which has its functional dependence on $B_{\mu\nu}$ through the form $B_{\mu\nu}B^{\mu\nu}$ \cite{Altschul:2009ae}. In our previous works on inflation with a single antisymmetric tensor field, we have considered potentials quadratic and quartic in the field $B_{\mu\nu}$ \cite{Aashish:2018lhv,Aashish:2021gdf}. In this study, we consider a potential term containing terms quadratic in each of the individual fields. Hence $V=B^{(k)}_{\mu\nu}B^{(k)\mu\nu}$. With all this initial setup, we find Einstein equations by varying the action eq. (\ref{2.1}) with respect to the metric tensor $g_{\mu\nu}$. The Einstein equations in tensor form is obtained as, 
\begin{equation} \label{2.2}
     G_{\mu\nu}=\kappa T_{\mu\nu}
\end{equation}
where, $G_{\mu\nu}$ is the Einstein tensor which comes from the Einstein-Hilbert part of our action. $T_{\mu\nu}$ is the stress-energy tensor obtained by varying the residual part of the action with respect to the metric tensor $g_{\mu\nu}$. For a flat FLRW background spacetime, the components of $G_{\mu\nu}$ reads as,
\begin{equation} \label{2.3}
    G_{00}=3H^2, \hspace{1em} G_{0i}=0, \hspace{1em} G_{ij}=-a^2(2\dot{H}+3H^2)     
\end{equation}
Here, $a(t)$ is the scale factor of expansion and $H={\dot{a}(t)}/{a(t)}$ is the Hubble parameter.  The components of $T^{M}_{\mu\nu}$ are given as,
\begin{equation} \label{2.4}
    T^M_{00}=\frac{3\Dot{B}^2+m^2B^2}{2a^4}, \hspace{1em} T^M_{0i}=0, \hspace{1em} T^M_{ij}=\frac{-\Dot{B}^2+m^2B^2}{2a^4}\delta_{ij}
\end{equation}
The stress energy tensor is diagonal for this ansatz of triplet fields. We can redefine $B(t)$ as $B(t)=a(t)^2\phi(t)$ to simplify our equations \cite{Aashish:2018lhv}. The equations are now,
\begin{equation} \label{2.5}
    H^2=\frac{\kappa}{2}[(\Dot{\phi}+2H\phi)^2+m^2\phi^2]
\end{equation}
\begin{equation} \label{2.6}
    2\Dot{H}+3H^2=\frac{\kappa}{2}[(\Dot{\phi}+2H\phi)^2-m^2\phi^2]
\end{equation}
Unlike the single field models, here we don't need to impose any constraints on the value of $B(t)$ to ensure a diagonal energy momentum tensor. Thus our theory can be considered less constrained compared to the predecessor antisymmetric tensor field inflation models \cite{Aashish:2018lhv,Aashish:2019zsy,Aashish:2020mlw,Aashish:2021gdf}.
\subsection{The de-Sitter solutions}
First, we will check whether our model permits de-Sitter solutions. The spacetime will begin to evolve from the de-Sitter solutions. During inflationary epoch, the Hubble parameter $H$ can be considered a constant due to the exponential expansion of the universe. Further, the field $\phi$ goes through the slow rolling phase after evolving from the de-Sitter value for about 70 e-folds. The value of $\phi$ does not vary significantly during the slow roll phase. Hence, $\phi$ can be treated as a constant while deducing the de-Sitter solutions. Looking for such solutions, we apply the constraint $\dot\phi\sim\dot H\sim 0$ into our system of equations, eq. (\ref{2.5}) and eq. (\ref{2.6}). We get the following solutions,
\begin{equation} \label{2.7}
    \phi_d^2=\frac{1}{\kappa} \hspace{2em} H_d^2=-\frac{m^2}{2}
\end{equation}
But this solution is inconsistent as can be seen from eq. (\ref{2.7}) that $H_d^2$ term is negative. So, we now try to modify our model by incorporating a non-minimal coupling term in our action eq. (\ref{2.1}).
\section{Non minimal coupling}\label{sec:3}
We have seen that the minimally coupled model cannot account for de-Sitter solutions. This motivates us to work with nonminimal models. Previous models on antisymmetric tensor field inflation employed non-minimal couplings for obtaining consistent de-Sitter solutions and tuning the primordial gravitational wave velocity to the velocity of light in vacuum \cite{Aashish:2018lhv,Aashish:2020mlw,Aashish:2021gdf}. These models have incorporated non-minimal couplings with Ricci scalar and Ricci tensor. The study on the single field inflation analogous to our choice of the triplet fields have shown that the primordial GW velocity can be tuned to $c$ in the presence of the coupling with Ricci tensor irrespective of the presence of the Ricci scalar coupling \cite{Aashish:2020mlw}. So we add the nonminimal coupling term between the tensor field triplet and Ricci tensor to our initial action eq. (\ref{2.1}). The strength of the coupling is controlled by the parameter $\zeta$ which has the dimensions of $M_{pl}^{-2}$. The action now becomes, 
\begin{multline}\label{3.1}
S=\int d^{4}x\sqrt{-g}[\frac{R}{2\kappa}-\frac{1}{12}H^{(k)}_{\lambda\mu\nu}H^{(k)\lambda\mu\nu}-B^{(k)}_{\mu\nu}B^{(k)\mu\nu}+\\ \frac{\tau}{2}(\nabla_{\lambda}B^{(k)\lambda\nu})(\nabla_{\mu}{B^{(k)\mu}}_{\nu})+\frac{\zeta}{2\kappa}B^{(k)\lambda\nu}{B^{(k)\mu}}_{\nu}R_{\lambda\mu}]
\end{multline}
The Einstein's equations now get modified into,
\begin{equation} \label{3.2}
    H^2+2\zeta H\phi\Dot{\phi}=\frac{\kappa}{2}[(\Dot{\phi}+2H\phi)^2+m^2\phi^2]
\end{equation}
\begin{multline} \label{3.3}
    2\Dot{H}+3H^2+\zeta(2\phi\Ddot{\phi}+2\Dot{\phi}^2-4\Dot{H}\phi^2-12H^2\phi^2)=\\ \frac{\kappa}{2}[(\Dot{\phi}+2H\phi)^2-m^2\phi^2]
\end{multline}
The de-Sitter Solutions are given as, 
\begin{equation}\label{3.4}
     \phi_d^2=\frac{1}{\kappa+3\zeta} \hspace{2em} H_d
     ^2=\frac{\kappa m^2}{6\zeta-2\kappa}
\end{equation}
 From now onwards we will be using the dimensionless parameter $y=\zeta/\kappa$ instead of the coupling strength $\zeta$. We can see here that $H_d^2$ can be made positive by constraining $y$. Positivity of the values found in eq. (\ref{3.4}) can be ensured by keeping $y>1/3$. These de-Sitter solutions should be stable to small fluctuations. We now check the stability by perturbing the Einstein equations around the de-Sitter background. We substitute $\phi$ as $\phi_d+\delta\phi$ and $H$ as $H_d+\delta H$ in equations (\ref{3.2}) and (\ref{3.3}). At linear order, the perturbed equations can be cast into a matrix form given as,
\begin{equation}\label{3.5}
    \dot{\Theta}=\Lambda\Theta
\end{equation}
with $ \Theta=\begin{pmatrix}
            \delta\phi \\
            \delta H
            \end{pmatrix}
            $, and the coefficient matrix $\Lambda$ as, 
\begin{equation} \label{3.6}
   \Lambda=\begin{pmatrix}
            -\frac{\kappa(m^2+4H_d^2)}{2H_d(\zeta+\kappa)} & \frac{1-2\kappa\phi_d^2}{\phi_d(\zeta+\kappa)} \\
            \frac{\phi_d(\kappa m^2(\zeta+2\kappa)-4\zeta(6\zeta+7\kappa)H_d^2)}{2(\zeta+\kappa)(-1+2\zeta\phi_d^2)} & \frac{H_d(3\zeta+2\kappa-2\zeta(6\zeta+7\kappa)\phi_d^2)}{(\zeta+\kappa)(-1+2\zeta\phi_d^2)}
    \end{pmatrix}
\end{equation}
The solution of this differential equation is given as,
\begin{equation}\label{3.7}
    \Theta=C_1e^{\lambda_1t}+C_2e^{\lambda_2t}
\end{equation}
where $\lambda_1$ and $\lambda_2$ are the eigen values of the coefficient matrix $\Lambda$. $C_1$ and $C_2$ are column vectors consisting of arbitrary integration constants. We can calculate the eigen values from the trace and determinant of $\Lambda$,
\begin{equation}\label{3.8}
    \lambda_1+\lambda_2=Tr(\Lambda)=\alpha H_d
\end{equation}
\begin{equation}\label{3.9}
    \lambda_1\lambda_2=Det(\Lambda)=\beta H_d^2
\end{equation}
with,
\begin{equation}\label{3.10}
    \alpha=\frac{\kappa(\zeta-3\kappa)}{(\zeta+\kappa)^2}=\frac{y-3}{(1+y)^2}, \hspace{0.5em} \beta=\frac{4(-9\zeta^2+\kappa^2)}{(\zeta+\kappa)^2}=\frac{4-36y^2}{(1+y)^2}
\end{equation}
Thus the eigen values are found to be,
\begin{equation}\label{3.11}
    \lambda_{1(2)}=\frac{H_d}{2}(\alpha+(-)\sqrt{\alpha^2-4\beta})
\end{equation}
$\lambda_1$ is always positive and $\lambda_2$ is always negative for our parameter range, i.e for $y>1/3$. If $C_1$ in eq. (\ref{3.7}) becomes 0, then we will have a decaying solution. Thus the perturbations around the de-Sitter values will decay in time and the de-Sitter background will be stable. Though, it is not clear at this time how to obtain such a solution without constraining the parameter $y$ and the coefficients $(C_1,C_2)$, we leave this problem for consideration in the future.

\section{Slow roll analysis}\label{sec:4}
The de-Sitter solutions correspond to an idealistic scenario where $H$ and $\phi$ are strictly constants. We now relax the strictness on the constancy of $H$ and $\phi$. This is called quasi-de-Sitter or slow roll limit. For addressing a viable inflationary scenario, the model should be able to support at least 70 e-folds of inflation \cite{Dodelson:2003ft}. This condition can be accomplished through the slow roll parameters.
\subsection{Slow roll parameters}
We can choose the same slow roll parameters used in the predecessor models \cite{Aashish:2020mlw,Aashish:2021gdf}. They are defined as,
\begin{equation}\label{4.1}
    \epsilon=-\frac{\dot{H}}{H^2}, \hspace{2em} \delta=\frac{\dot{\phi}}{H\phi}
\end{equation}
Here, the slow roll parameter $\epsilon$ controls the acceleration of the universe. The parameter $\delta$ is related to the flatness of our potential $V(\phi)$. These slow roll parameters constitute of only first order time  derivatives whereas the popular scalar field inflation models constitute of second order time derivatives too. Having a complicated background structure, during slow roll analysis, we are looking for a simpler case where the slow roll parameters constitute only first order time derivatives.
We now differentiate Einstein equation, eq. (\ref{3.2}) with respect to time $t$. It can be expressed in terms of slow roll parameters as,
\begin{equation}\label{4.2}
    2H^2(\epsilon+2\kappa\phi^2(\delta-\epsilon))+m^2\kappa\phi^2\delta+O(\epsilon^2,\delta^2)=0
\end{equation}
Einstein equation eq. (\ref{3.3}) can be written as,
\begin{equation}\label{4.3}
    \epsilon(1-2\zeta\phi^2)+\kappa\phi^2\delta+O(\epsilon^2,\delta^2)=0
\end{equation}
From the above equations, we can obtain the following relation, up to linear order in slow roll parameters.
\begin{equation}\label{4.4}
    \delta=-\frac{2\epsilon(y-1)}{3 y}
\end{equation}
\begin{figure}[h!]
    \centering
    \includegraphics[width=1\linewidth,height=.8\linewidth]{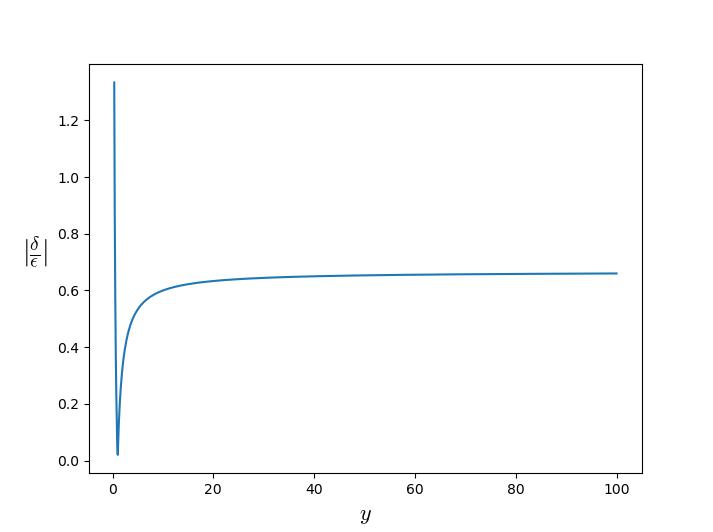}
    \caption{The figure shows the ratio of the absolute value of the slow roll parameters}
    \label{fig:1}
\end{figure}
\\
From fig. (\ref{fig:1}), we can see that the slow roll parameters are nearly of the same order for most of the $y$ range. A small value of $\delta$ will keep the driving potential $V(\phi)$ sufficiently flat. Also, this will ensure that $\epsilon$ is small, satisfying the requirements for slow roll inflation. The number of e-folds of inflation, $N$, can be related to the slow roll parameter $\delta$ by the following relation, 
\begin{equation} \label{4.5}
    N=\int^{t_f}_{t_i}Hdt=\int^{\phi_f}_{\phi_i}\frac{H}{\dot{\phi}}d\phi=\frac{1}{\delta}\int^{\phi_f}_{\phi_i}\frac{1}{\phi}d\phi=\frac{1}{\delta}ln\left(\frac{\phi_f}{\phi_i}\right)
\end{equation}
Here, the slow roll parameter $\delta$ appears in the denominator. Hence, by making $\delta$ small, enough number of e-folds can be ensured during inflation. We can write, 
\begin{equation} \label{4.6}
    \delta<\frac{1}{70}ln\left(\frac{\phi_f}{\phi_i}\right)
\end{equation}

\subsection{Evolution of slow roll parameters}
Now we look at the evolution of the slow roll parameters. For developing their dynamics, we need to rewrite our system of Einstein equations in terms of these slow roll parameters. We try to analyse the evolution with respect to the number of e-folds, $N$. The derivative with respect to $N$ is represented as a $'$ over the quantity. Differentiating the Einstein equations with respect to $t$ and rewriting them in terms of the slow roll parameters will yield,
\begin{dmath}\label{4.7}
    \epsilon\delta'-\epsilon^2\delta+2(-1+y)(\delta'+\epsilon^2)+(-1+4y)\delta^2+(1+2y)\epsilon\delta -2(1+3y)\delta+2(1-3y)\epsilon=0
\end{dmath}
\begin{dmath}\label{4.8}
    2\epsilon^3\delta-\epsilon'\epsilon\delta-2\epsilon^2\delta'+\epsilon'\delta'+(-2+8y)\delta^3+2(1-6y)\epsilon\delta^2+4(1-y)\epsilon^3-(1+4y)\epsilon^2\delta-(2-12y)\delta\delta' (1-2y)\epsilon'\delta+(3-6y)\epsilon\delta'+2(-1+y)\epsilon\epsilon'+2y\delta''-7\delta^2+3(1+6y)\epsilon\delta+2(-1+5y)\epsilon^2-2\delta'-2(1+y)\epsilon'+ 2(1-9y)\delta+2(-5+3y)\epsilon=0
\end{dmath}
where we have used the relations,
\begin{equation}\label{4.9}
    \frac{\dddot \phi}{H^3\phi}=\delta''-\epsilon\delta'-\epsilon'\delta+\frac{\Ddot{\phi}}{H^2\phi}(3\delta-2\epsilon)+2\epsilon\delta^2-2\delta^3
\end{equation}
\begin{equation}\label{4.10}
    \frac{1}{\kappa\phi^2}=1+3y-\frac{\Ddot{\phi}}{2H^2\phi}+\frac{\delta^2}{2}+\delta\left(\frac{1}{2}-2y \right)+\epsilon(1-y)
\end{equation}
\begin{equation}\label{4.11}
    \frac{\Ddot{\phi}}{H^2\phi}=\delta'+\delta^2-\epsilon\delta
\end{equation}
\begin{equation}\label{4.12}
    \frac{\Ddot{H}}{H^3}=2\epsilon^2-\epsilon'
\end{equation}
Differentiating eq. (\ref{4.7}) with respect to $N$, we get,
\begin{multline}\label{4.13}
    \epsilon'\delta'+\epsilon\delta''-2\epsilon\epsilon'\delta-\epsilon^2\delta'+2(-1+y)(\delta''+2\epsilon\epsilon')+2(-1+4y)\delta\delta'\\+(1+2y)(\epsilon'\delta+\epsilon\delta')-2(1+3y)\delta' +2(1-3y)\epsilon'=0
\end{multline}
\begin{figure*}[h!]
    \centering
    \begin{subfigure}{.5\textwidth}
    \centering
    \includegraphics[width=1.1\linewidth,height=.75\linewidth]{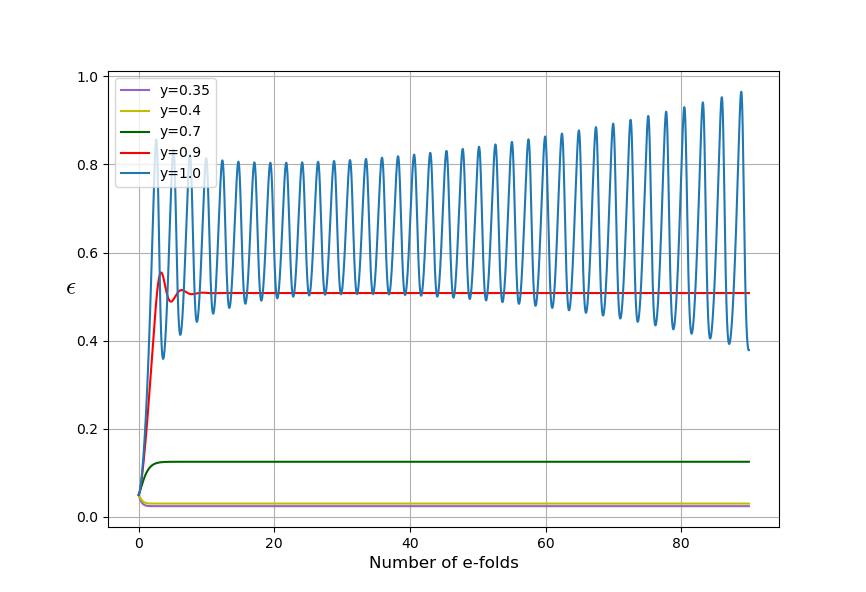}
    \end{subfigure}%
    \begin{subfigure}{.5\textwidth}
    \centering
    \includegraphics[width=1.1\linewidth,height=.75\linewidth]{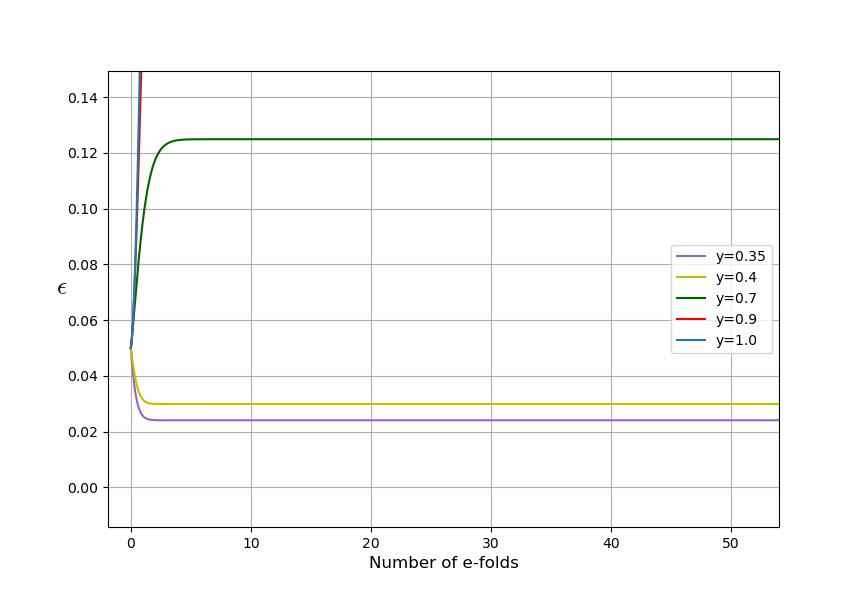}
    \end{subfigure}
    \caption{The figures show the evolution of $\epsilon$ with respect to $N$. The figure on right gives the enlarged view.}
    \label{fig:2}
\end{figure*}
\newpage
Eqs. (\ref{4.7}) and (\ref{4.13}) represent our system of equations which we try to evaluate using numerical integration. For numerical evaluation, we can define the variable,
\begin{equation}\label{4.14}
    \alpha=\delta' \hspace{1em} \Longrightarrow \hspace{1em} \alpha'=\delta''
\end{equation}
Thus the system of equations takes the form,
\begin{equation}\label{4.15}
    \alpha'=f_1(\alpha, \epsilon, \delta), \hspace{1em} \epsilon'=f_2(\alpha, \epsilon, \delta), \hspace{1em} \delta'=\alpha 
\end{equation}
with
\begin{equation}\label{4.16}
    f_1(\alpha, \epsilon, \delta)=\frac{f_{1n}(\alpha, \epsilon, \delta)}{f_{1d}(\alpha, \epsilon, \delta)}, \hspace{1em} f_2(\alpha, \epsilon, \delta)=\frac{f_{2n}(\alpha, \epsilon, \delta)}{f_{2d}(\alpha, \epsilon, \delta)}
\end{equation}

The explicit forms of the functions $f_{1n}, f_{2n}, f_{1d}$ and $f_{2d}$ are given in Appendix \ref{app:A}. Using these functional forms, we evaluate the system of equations given in eq. (\ref{4.15}) using 4th order Runge Kutta method. The system is evolved from an initial value of the order of $10^{-2}$ for the slow roll parameters. The results for different values of $y$ are shown in fig. (\ref{fig:2}).

We can see that as the value of $y$ increases the slow roll phase is short lived and the value of slow roll parameter $\epsilon$ increase rapidly such that it becomes significantly large. We have to choose the coupling values $y$ such that the slow roll phase is ensured for at least 70 e-folds of inflation. Since $y$ is constrained to be less than $1/3$, we can select a value of $y$ in the allowed range near $1/3$.

\section{Energy Constraints}\label{sec:5}
In this section, we look at the energy conditions satisfied by our tensor field theory. An energy condition, crudely speaking, is a relation one demands the stress-energy tensor to satisfy for invoking the notion that 'energy should be positive'. They are not exactly physical constraints for a system, but are rather mathematically imposed boundary conditions. The general energy conditions featured in general relativity are the Weak, Null, Strong and Dominant energy conditions. To put the energy conditions in a concrete form, we can assume the stress-energy tensor to admit the following form for our Friedmann background,
\begin{equation}\label{5.1}
    T^{\mu\nu}=\begin{bmatrix}
        \rho & 0 & 0 & 0 \\
        0 & \frac{p_1}{a^2} & 0 & 0 \\
        0 & 0 & \frac{p_2}{a^2} & 0\\
        0 & 0 & 0 & \frac{p_3}{a^2}
    \end{bmatrix}
\end{equation}
where, $\rho$ is the energy density and $p_i$ are the pressures. In our case, we can write,
\begin{equation}\label{5.2}
    \rho=\frac{3}{2}\left(m^2\phi^2+(\Dot{\phi}+2H\phi)^2-4yH\phi\Dot{\phi} \right)
\end{equation}
Similarly the pressure $p_i$ can be written as,
\begin{multline}\label{5.3}
    p_i=\frac{1}{2}[m^2\phi^2-(\Dot{\phi}+2H\phi)^2+2y(\phi\Ddot{\phi}+\Dot{\phi}^2-2\Dot{H}\phi^2\\-6H^2\phi^2) ]
\end{multline}
The requirements for the energy conditions are summarized in Table (\ref{table:1}).
\begin{table*}[h!]
\centering
\begin{tabular}{ p{4cm} p{5cm} p{4cm} }
\hline
\hline
Name & Statement & Conditions\\
\hline 
 Weak & $T_{\alpha\beta}v^{\alpha}v^{\beta} \geq 0$ & $\rho\geq 0$, $\rho+p_i>0$ \\  [0.5ex] 
 Null & $T_{\alpha\beta}k^{\alpha}k^{\beta} \geq 0$ & $\rho+p_i\geq0$ \\   [0.5ex] 
 Strong & ($T_{\alpha\beta}-\frac{1}{2}Tg_{\alpha\beta})v^{\alpha}v^{\beta} \geq 0$ & $\rho+{\sum}_ip_i\geq 0$,  $\rho+p_i\geq0$ \\ [0.5ex] 
 Dominant & $-{T^{\alpha}}_{\beta}v^{\beta}$ future directed & $\rho\geq 0$, $\rho\geq|p_i|$
\end{tabular}
\caption{Energy Conditions}
\label{table:1}
\end{table*}
\\
For our model, we can write the following relations upto first order in slow roll parameters,
\begin{equation}\label{5.4}
    \frac{1}{\kappa\phi^2}\approx1+3y+\frac{\epsilon}{2}(1+y)+\delta\left(1-\frac{y}{2} \right)+O(\epsilon^2,\delta^2,\epsilon\delta)
\end{equation}
\begin{equation}\label{5.5}
    \frac{m^2}{2H^2}\approx -1+3y+\frac{\epsilon}{2}(1+y)+\delta\left(-1+\frac{3y}{2} \right)+O(\epsilon^2,\delta^2,\epsilon\delta)
\end{equation}
Using the above relations, we can approximate the energy density and pressure as,
\begin{equation}\label{5.6}
    \rho\approx \frac{3}{2}H^2\phi^2(2+6y+\epsilon(1+y)+\delta(2-y))
\end{equation}
\begin{equation}\label{5.7}
    p_i\approx -H^2\phi^2\left(3+9y-\frac{\epsilon}{2}(1+9y)+\delta\left(3-\frac{3y}{2} \right) \right)
\end{equation}

From the form of $\rho$ and $p_i$ given in eqs. (\ref{5.6}) and (\ref{5.7}), it is straightforward to see that Weak, Null and Dominant energy conditions are satisfied, whereas the Strong energy condition is violated. This is a typical feature of cosmic inflation models where the positive acceleration of the universe requires $\rho+3p$ to be negative \cite{Visser:1995cc}.

\section{Perturbations}\label{sec:6}
We now look at the possible perturbations in our model. The perturbations can simultaneously arise from both the metric as well as from the triplet of fields driving inflation. Following SVT decomposition, \cite{Dodelson:2003ft,Guzzetti:2016mkm}, the metric perturbations can be expressed in the following form,
\begin{equation} \label{6.1}
\begin{array}{c}
    \delta g_{00}=-\psi \hspace{1cm} \delta g_{0i}=a(\partial_i \chi +E_i) \\
    \vspace{0.1mm}\\
    \delta g_{ij}=a^2(-2\alpha\delta_{ij}+2\partial_{ij}\beta+(\partial_iF_j+\partial_jF_i)+h_{ij})
    \end{array}
\end{equation}
where $\psi$,$\chi$,$\alpha$ and $\beta$ are the scalar modes, $E_i$ and $F_i$ are divergence free vector modes, and $h_{ij}$ constitutes the traceless, non-transverse tensor modes. Following the SVT decomposition theorem, the dynamical equations for scalar, tensor and vector modes separate at linear order and therefore can be studied individually. Initially, we are switching off the scalar and vector modes and considering only the tensor perturbations coming from the metric. The perturbations coming from the field triplet take the following form,
\begin{equation}\label{6.2}
    \begin{array}{c}
        \delta B^{(1)}_{0i}=-E_i, \hspace{2em} \delta B^{(1)}_{0i}=\epsilon_{ijk}M_k   \\
         \\
        \delta B^{(2)}_{0i}=-F_i, \hspace{2em} \delta B^{(2)}_{0i}=\epsilon_{ijk}N_k   \\
        \\
         \delta B^{(3)}_{0i}=-G_i, \hspace{2em} \delta B^{(3)}_{0i}=\epsilon_{ijk}O_k
    \end{array}
\end{equation}
Consider the perturbations from the first field $\delta B^{(1)}_{\mu\nu}$. They can be decomposed in the following manner.
\begin{equation} \label{6.3}
    \Vec{E}=\Vec{\nabla}u+\Vec{U}, \hspace{1cm} \Vec{M}=\Vec{\nabla}v+\Vec{V}
\end{equation}
where $\Vec{U}$ and $\Vec{V}$ are divergence free vector fields. Similar decomposition holds for the vectors $\Vec{F}, \Vec{N}, \Vec{G}$ and $\Vec{O}$ respectively.  It is evident from the above that such a structure doesn't permit tensor modes. Hence, the perturbations from the field triplet do not couple with the tensor perturbations coming from the metric, and hence they can be studied separately. Initially, we look at the field triplet perturbations, keeping our metric at the background value. We follow the perturbative analysis performed in Refs. \cite{Aashish:2019zsy,Aashish:2021gdf} to check for ghost instabilities. Ghost instabilities are seen when the coefficients of the kinetic terms in the perturbed action acquire negative coefficients. They make the theories ill defined and makes the energy to be unbounded from below. Further, we make use of Fourier space, to get rid of the spatial derivatives in our perturbed action. The coefficient matrix is given by,
\begin{equation}\label{6.4}
    T=diag\begin{bmatrix}
    \underbrace{\frac{k^2}{2a(t)} \ \frac{1}{2a(t)} \ \frac{1}{2a(t)} \  \frac{k^2a(t)\tau}{2} \ \frac{a(t)\tau}{2} \ \frac{a(t)\tau}{2} }_{\text{$3$~times }}  & \cdots & \cdots
    \end{bmatrix}
\end{equation}
The coefficient matrix consists of the above block repeated thrice. Since our coefficient matrix is diagonal, the diagonal elements constitute the eigen values. They are positive, given the condition $\tau>0$. Thus, our theory is free from ghost instabilities. We note that these ghost-free conditions are subject to the conditions akin to the results of Refs. \cite{Aashish:2019zsy,Aashish:2021gdf}, particularly our choice of the background fields $B_{\mu\nu}^{(j)}$, the metric, and considering only the perturbations in the background field triplet. In a more general scenario, we can expect pathological instabilities since all the eighteen degrees of freedom are propagating, and will be the subject of future endeavors. Also, the general analysis would involve the scalar and vector modes in the metric as well and will be addressed in future works as the analysis will be cumbersome. Now, we look at the other part, consisting of the tensor modes from the metric. Here, we keep the field triplet at their background values. The metric tensor in this scenario reads as,
\begin{equation} \label{6.5} 
    g_{00}=-1, \hspace{1cm} g_{0i}=0, \hspace{1cm} g_{ij}=a^2(\delta_{ij}+h_{ij})
\end{equation}
The tensor perturbation $h_{ij}$ is inherently gauge invariant at linear order \cite{Malik:2008im}. These tensor modes are expected to propagate as gravitational waves. The presence of such primordial gravitational waves have been hypothesized in several models of inflation \cite{Starobinsky:1979ty,Abbott:1984fp,Rubakov:1982df,Fabbri:1983us}. These primordial GWs are of great interest as they are expected to have signatures regarding the early universe physics. We choose the following form for $h_{ij}$ which satisfies its non-transverse nature and vanishing trace, \cite{Dodelson:2003ft},
\begin{equation} \label{6.6}
    h_{ij}=\begin{pmatrix}
    h_+ & h_{\times} & 0 \\
    h_{\times} & -h_+ & 0\\
    0 & 0 & 0 
    \end{pmatrix}
\end{equation}
Here $h_+$ and $h_{\times}$ correspond to the two GW polarizations. In this configuration, they lie in the $X-Y$ plane, whereas the wave vector $\Vec{k}$ is oriented in $Z$ direction. With all this information at our disposal, we apply them to our non-minimally coupled action. We analyse the action at second order in the perturbations. This is because the kinetic terms arise at second order in perturbations. Also, the second order action is expected to give rise to gravitational waves \cite{Guzzetti:2016mkm}. The second order action can be expressed as,
\begin{dmath} \label{6.7}
    S_2^{FT}=\sum_{e=+,\times}\int dtd^3k\frac{a^3}{4\kappa}[\Omega_k\dot{h_e}^{\dagger}\dot{h_e}+\Omega_c(\dot{h_e}^{\dagger}h_e+h_e^{\dagger}\dot{h_e})+\Omega_gh_e^{\dagger}h_e]
\end{dmath}
The explicit form of the functions in this expression is given as,
\begin{equation}\label{6.8}
    \Omega_k=1+2\zeta\phi^2
\end{equation}
\begin{equation}\label{6.9}
    \Omega_c=-2H+6\zeta(\phi\Dot{\phi}-H\phi^2)
\end{equation}
\begin{dmath}\label{6.10}   
\Omega_g=6\zeta(\Dot{H}\phi^2+3H^2\phi^2) +3\kappa((\Dot{\phi}+2H\phi)^2-m^2\phi^2)-6(\Dot{H}+2H^2)-\frac{k^2}{a^2}(1+(4\zeta-4\kappa\tau)\phi^2)
\end{dmath}
For eliminating ghost instabilities, the $\Omega_k$ here must be positive.
In the slow roll limit, we can approximate,
\begin{equation}\label{6.11}
    \frac{1}{\kappa\phi^2}\approx1+3y+\frac{\epsilon}{2}(1+y)+\delta\left(1-\frac{y}{2} \right)+O(\epsilon^2,\delta^2,\epsilon\delta)
\end{equation}
Using this approximation in our expression for $\Omega_k$ gives,
\begin{equation}\label{6.12}
    \Omega_k\approx\frac{1+5y}{1+3y}+\left(\frac{y}{1+3y^2} \right)(\delta(-2+y)-\epsilon(1+y))
\end{equation}
In the above equation, $\delta$ can be related to $\epsilon$ through eq. (\ref{4.4}). $\epsilon$ takes values between 0 and 1. The smallness of the slow roll parameters will make the expression for $\Omega_k$ dominated by its de-Sitter part. Thus for a $y$ value near $1/3$, $\Omega_k$ is positive. Hence, we can see that ghost instabilities are absent. The tensor perturbations propagate in the form of gravitational waves. Initially, we perform an estimate of the GW velocity. Varying the action, eq. (\ref{6.7}), with respect to $h_e^{\dagger}$ we can obtain the equation of motion.
\begin{equation}\label{6.13} 
    \ddot{h_e}+\left(\frac{\dot{\Omega_k}}{\Omega_k}+3H\right)\dot{h_e}+\left(\frac{\dot{\Omega_c}+3H\Omega_c-\Omega_g}{\Omega_k}\right)h_e=0
\end{equation}
In the slow roll limit, using eq. (\ref{6.11}), we can estimate the functions involved in the equation motion.
\begin{equation}\label{6.14}
    \frac{\dot{\Omega_k}}{\Omega_k}=\frac{4y\kappa\phi^2}{1+2y\kappa\phi^2}H\delta\approx \frac{4y}{1+5y}H\delta
\end{equation}
\begin{equation}\label{6.15}
    \Dot{\Omega_c}+3H\Omega_c-\Omega_g=F(H,\phi)+\frac{k^2}{a^2}(1+(4\zeta-4\kappa\tau)\phi^2)
\end{equation}
where the function $F$ is given as,
\begin{dmath}\label{6.16}    F(H,\phi)=4\Dot{H}+6H^2+3\kappa(m^2\phi^2-(\Dot{\phi}+2H\phi)^2)+6\zeta(\phi\Ddot{\phi}+\Dot{\phi}^2-2\Dot{H}\phi^2+H\phi\Dot{\phi}-6H^2\phi^2)
\end{dmath}
Using Einstein equations, we can write,
\begin{equation}\label{6.17}
    F(H,\phi)=-2\dot{H}-6H^2+3\kappa m^2\phi^2
\end{equation}
We are placing an additional constraint on the value of $m$, which is $m=(\frac{\Dot{\phi}}{\phi}+2H)$. Why we employ such a condition will be discussed in the end. Now in this limit,
\begin{equation}\label{6.18}
    F(H,\phi)=H^2(2\epsilon-6+\kappa\phi^2(4+4\delta))
\end{equation}
Thus, we have,
\begin{equation}\label{6.19}
    \frac{\dot{\Omega_c}+3H\Omega_c-\Omega_g}{\Omega_k}=QH^2+\frac{k^2}{a^2}P
\end{equation}
where the functions are given as, 
\begin{equation}\label{6.20}
    Q=\frac{2\epsilon-6+\kappa\phi^2(4+4\delta)}{1+2y\kappa\phi^2} \hspace{2em} P=\frac{1+\kappa\phi^2(4y-4\tau)}{1+2y\kappa\phi^2} 
\end{equation}
In the quasi de-Sitter limit, we can approximate,
\begin{equation}\label{6.21}
    Q\approx \frac{6-18y}{1+5y}+4\epsilon\left(\frac{-1+y+6y^2}{(1+5y)^2} \right)+6\delta\left(\frac{9y+y^2}{(1+5y)^2} \right) 
\end{equation}
\begin{equation}\label{6.22}
    P\approx 1+\frac{y-2\tau}{1+5y}\left(2-\epsilon\left(\frac{1+y}{1+5y} \right)+\delta\left( \frac{y-2}{1+5y}\right) \right)
\end{equation}
We substitute a wave solution in eq. (\ref{6.13}) of the form  $h_e=A\exp[-i\int^t(c_Tk/a(t'))dt']\Vec{e}$ . Here $c_T$ is the velocity of propagation of the gravitational wave, $A$ is a constant and $\Vec{e}$ is a constant vector. The dispersion relation turns out to be,
\begin{equation}\label{6.23} 
    c_T^2+i\left(\frac{2aH}{k}\right)\left(1+\frac{2y}{1+5y} \right)c_T-\left(\frac{aH}{k}\right)^2Q-P=0
\end{equation}
For the deep subhorizon case, we have $k>>aH$. In this limit, we can neglect the terms proportional $\frac{aH}{k}$. So,
\begin{equation}\label{6.24} 
    c_T^2=P\approx1+\frac{y-2\tau}{1+5y}\left(2-\epsilon\left(\frac{1+y}{1+5y} \right)+\delta\left( \frac{y-2}{1+5y}\right) \right)
\end{equation}
In eq. (\ref{6.24}), for $y=2\tau$, we have $c_T^2=1$. Hence, gravitational waves propagate with the velocity of light in vacuum. This value of GW velocity is consistent with the recent observations of GWs originating from astrophysical sources. But a solid proof can only be obtained after their detection because unlike those from astrophysical sources, they are of primordial origin. Actually, there is no stringent constraint from inflation on primordial gravitational wave velocity. But the equality of GW velocity with $c$ can be necessitated if we are to treat $B$ fields as dark matter or dark energy.

Now we try to solve eq. (\ref{6.13}). We adopt the analysis performed in Refs. \cite{Dodelson:2003ft,Aashish:2021gdf}.The temporal dependence of the equations from now will be expressed in terms of the conformal time coordinate $\eta$. The derivative with respect to $\eta$ will be denoted by a prime over the quantity.  Substituting the explicit forms of the coefficients in eq. (\ref{6.13}), we have,

\begin{figure*}[h!]
    \centering
    \begin{subfigure}{.5\textwidth}
    \centering
    \includegraphics[width=1\linewidth,height=.8\linewidth]{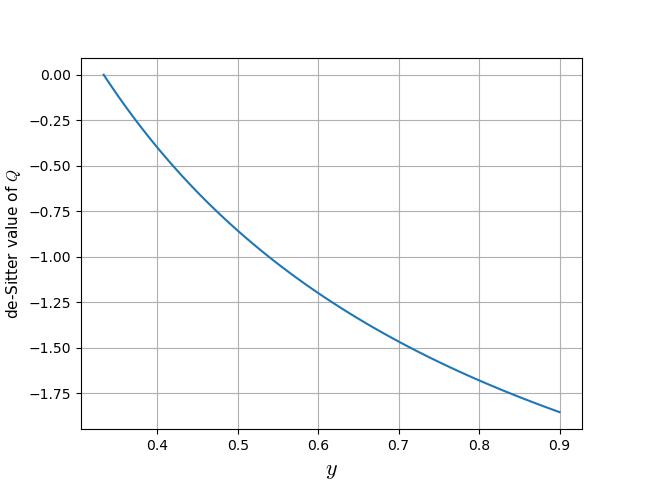}
    \end{subfigure}%
    \begin{subfigure}{.5\textwidth}
    \centering
    \includegraphics[width=1\linewidth,height=.8\linewidth]{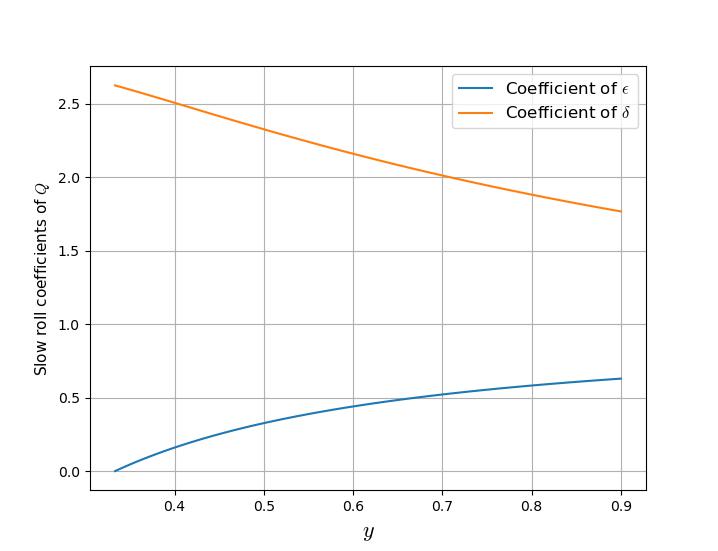}
    \end{subfigure}
    \caption{The figure shows the estimate of the functions involved in the expression for $Q$}
    \label{fig:3}
\end{figure*}

\begin{equation} \label{6.25}
    h_e''+\left(2+\frac{4y}{1+5y}\delta \right)aHh_e'+[k^2+a^2H^2Q]h_e=0
\end{equation}
where, we have approximated $P$ as 1. Using the transformation, $h_e=a^{-\lambda}\Tilde{h_e}$, where $\lambda=1+\frac{2y}{1+5y}\delta$, we can rewrite the above equation as,
\begin{equation} \label{6.9}
\Tilde{h_e}''+\Tilde{h_e}\left(k^2+a^2H^2\left( 1+\epsilon-3\lambda+Q\right) \right)=0
\end{equation}
We can split the function $Q$ into a de-Sitter part, $Q^{dS}$, which is the part of $Q$ independent of slow roll parameters(we can obtain this by setting $\epsilon$ and $\delta$ to 0 which is the condition we used to obtain de-Sitter solutions), and two other parts, $Q^{\epsilon}$ and $Q^{\delta}$, which are the respective coefficients of the slow roll parameters $\epsilon$ and $\delta$ in our expression for $Q$ given in eq. (\ref{6.21}). Note that we will be employing similar terminologies for the quantities which we will encounter. The estimate of these functions is shown in fig. (\ref{fig:3}). Thus,
\begin{multline}\label{6.27}
     \Tilde{h_e}''+\Tilde{h_e}[k^2+a^2H^2( 1-3\lambda+Q^{dS}+\epsilon(1+Q^{\epsilon} ) \\ +Q^{\delta}\delta) ]=0
\end{multline}
In the slow roll limit, we have, $aH\approx-(1+\epsilon)/\eta$. Thus, we obtain,
\begin{equation}\label{6.28}
    \Tilde{h_e}''+\left(k^2-\frac{\omega^2}{\eta^2}\right)\Tilde{h_e}=0
\end{equation}
where, 
\begin{equation}\label{6.29}
    \omega^2=3\lambda-1-Q^{dS}+\epsilon\left(3-Q^{\epsilon}-2Q^{dS}\right)-Q^{\delta}\delta
\end{equation}
 Eq. (\ref{6.28}) corresponds to the equation of a harmonic oscillator. We now quantize this oscillator and decompose it in terms of the creation and annihilation operators.
\begin{equation}\label{6.30} 
    \hat{\Tilde{h_e}}(k,\eta)=\nu_e(k,\eta)\hat{a}_{\Vec{k}}+\nu_e(k,\eta)^*\hat{a}^{\dagger}_{\Vec{k}}
\end{equation}
where $\nu_e(k,\eta)$ should satisfy,
\begin{equation}\label{6.31} 
     \nu_e(k,\eta)''+\left(k^2-\frac{\omega^2}{\eta^2}\right)\nu_e(k,\eta)=0
\end{equation}
 We define the variables $p=-k\eta$ and $\Tilde{\nu_e}=p^{-\frac{1}{2}}\nu_e$. In terms of these variables eq. (\ref{6.31}) can be rewritten into a Bessel differential equation of order parameter $\nu$ as follows,
\begin{equation} \label{6.32}
    p^2\frac{d^2\Tilde{\nu_e}}{dp^2}+p\frac{d\Tilde{\nu_e}}{dp}+(p^2-\nu^2)\Tilde{\nu_e}=0
\end{equation}
Here $\nu^2=\omega^2+\frac{1}{4}$. The order parameter $\nu$ can be approximated as,
\begin{equation}\label{6.33}
    \nu\approx\frac{3p}{2}+\epsilon\left(\frac{1-q}{p} \right)+\delta\left(\frac{f-Q/3}{p} \right)
\end{equation}
with,
\begin{equation}\label{6.34}
    f=\frac{2y}{1+5y}, \hspace{2em} p=\sqrt{1-\frac{4}{9}Q^{dS}}, \hspace{2em} q=\frac{Q^{\epsilon}+2Q^{dS}}{3}
\end{equation}
The solution of Bessel's differential equation can be written as a sum of the Hankel functions of first and second kind. In our case, we get,
\begin{equation} \label{6.35}
    \nu_e(p)=\sqrt{p}[A_1H_{\nu}^{(1)}(p)+A_2H_{\nu}^{(2)}(p)]
\end{equation}
where $H_{\nu}^{(1)}(p)$ and $H_{\nu}^{(2)}(p)$ are Hankel functions of the first and second kind respectively. Note that subhorizon and super horizon limits are represented by the conditions $p>>1$ and $p<<1$ respectively. First we look at the subhorizon case. For $p>>1$, we have the following asymptotic forms for the Hankel functions \cite{Guzzetti:2016mkm,Aashish:2021gdf}, 
\begin{equation} \label{6.36}
    H_{\nu}^{(1)}(p)\approx \sqrt{\frac{2}{\pi p}}e^{-\frac{i\pi}{4}(1+2\nu)}e^{ip}
\end{equation}
\begin{equation} \label{6.37}
     H_{\nu}^{(2)}(p)\approx \sqrt{\frac{2}{\pi p}}e^{\frac{i\pi}{4}(1+2\nu)}e^{-ip}
\end{equation}
\begin{figure*}[h!]
    \centering
    \begin{subfigure}{.5\textwidth}
    \centering
    \includegraphics[width=1\linewidth,height=.8\linewidth]{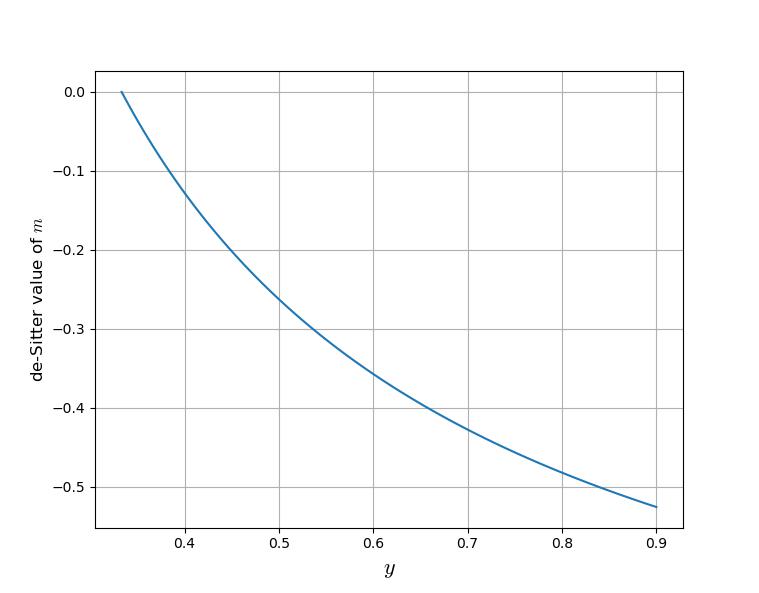}
    \end{subfigure}%
    \begin{subfigure}{.5\textwidth}
    \centering
    \includegraphics[width=1\linewidth,height=.8\linewidth]{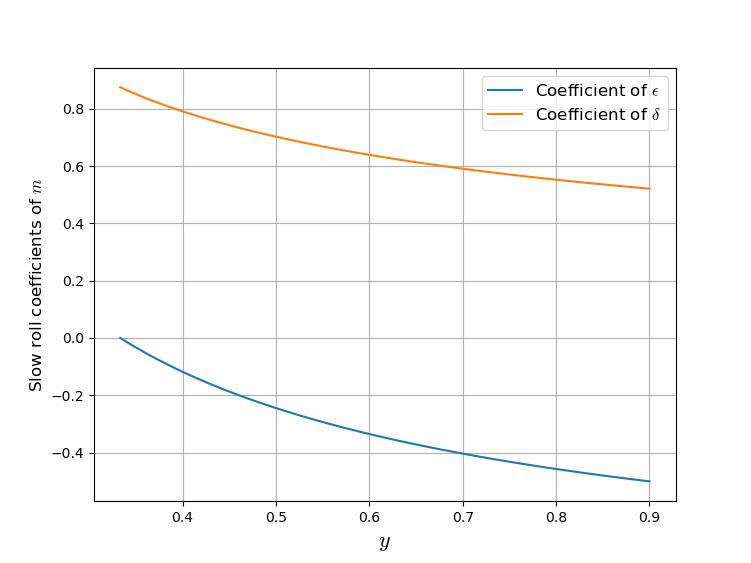}
    \end{subfigure}
    \caption{The figure shows the estimate of the functions involved in the expression for $m$}
    \label{fig:4}
\end{figure*}

We need to match our solution with boundary values to determine the coefficients $A_1$ and $A_2$. For this, we assume the universe at early times to be in the Bunch-Davies vacuum state \cite{Kundu:2011sg}.
The properly normalized solution will be,
\begin{equation} \label{6.38}
    \nu_e(k,\eta)=\frac{1}{\sqrt{2k}} e^{-ik\eta}
\end{equation}
Matching with this plane wave solution in the subhorizon limit, we have,
\begin{equation}\label{6.39}                      
    A_1=\frac{1}{2}\sqrt{\frac{\pi}{k}}e^{i\left(\nu+\frac{1}{2}\right)\frac{\pi}{2}}, \hspace{2em} A_2=0
\end{equation}
Thus the exact solution in the subhorizon limit will be,
\begin{equation} \label{6.40}
    \nu_e(k,\eta)=\frac{\sqrt{\pi}}{2}e^{i\left(\nu+\frac{1}{2}\right)\frac{\pi}{2}}\sqrt{-\eta} H_{\nu}^{(1)}(-k\eta)
\end{equation}
For $p<<1$, $H_{\nu}^{(1)}$ takes the following asymptotic form.
\begin{equation} \label{6.41}
    H_{\nu}^{(1)}(p)\approx\sqrt{\frac{2}{\pi}}\frac{\Gamma(\nu)}{\Gamma(3/2)}e^{-i\frac{\pi}{2}}2^{\nu-\frac{3}{2}}p^{-\nu}
\end{equation}
Thus in the super horizon limit we have,
\begin{equation} \label{6.42}
    \nu_e(k,\eta)=\frac{\Gamma(\nu)}{\Gamma(3/2)}e^{i\left(\nu-\frac{1}{2}\right)\frac{\pi}{2}}2^{\nu-\frac{3}{2}}\frac{1}{\sqrt{2k}}(-k\eta)^{\frac{1}{2}-\nu}
\end{equation}
where $\Gamma$ is the Euler function. $a$ varies as $\eta^{-(1+\epsilon)}$ in the quasi de-Sitter limit. Hence, the tensor modes become,
\begin{equation}\label{6.43}
    h_e(k,\eta)=\frac{\nu_e(k,\eta)}{a}=\frac{C}{k^{\lambda+\epsilon+\frac{1}{2}}}(-k\eta)^{\lambda+\epsilon+\frac{1}{2}-\nu}
\end{equation}
where $C$ is given by $\Gamma(\nu)2^{\nu-1}(-1)^{-\lambda-\epsilon}e^{i(\nu-\frac{1}{2})\frac{\pi}{2}}/\sqrt{\pi}$.
Thus in the superhorizon limit, we have,
\begin{equation}\label{6.44}
    h_e\propto (-\eta)^m
\end{equation}
where $m$ is given by,
\begin{equation}\label{6.45}
m=\frac{3}{2}(1-p)+\epsilon\left(1+\frac{q-1}{p} \right)+\delta\left(f+\frac{Q^{\delta}-3f}{3p} \right)
\end{equation}
$p$, $q$ and $f$ are the functions defined in eq. (\ref{6.34}). The dependence of the tensor mode amplitude on super horizon scales depends on the parameter $m$. The estimate of $m$ in the slow roll scenario is depicted in fig. (\ref{fig:4}). 

\begin{figure*}[h!]
    \centering
    \begin{subfigure}{.5\textwidth}
    \centering
    \includegraphics[width=1\linewidth,height=.8\linewidth]{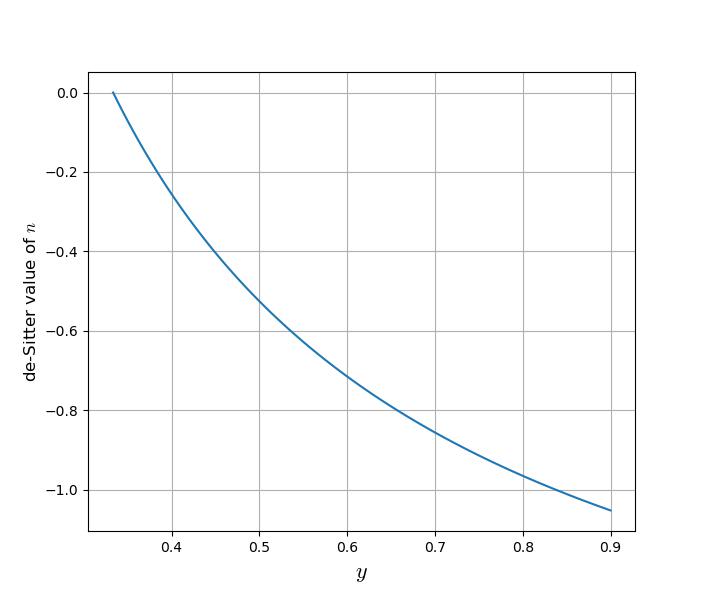}
    \end{subfigure}%
    \begin{subfigure}{.5\textwidth}
    \centering
    \includegraphics[width=1\linewidth,height=.8\linewidth]{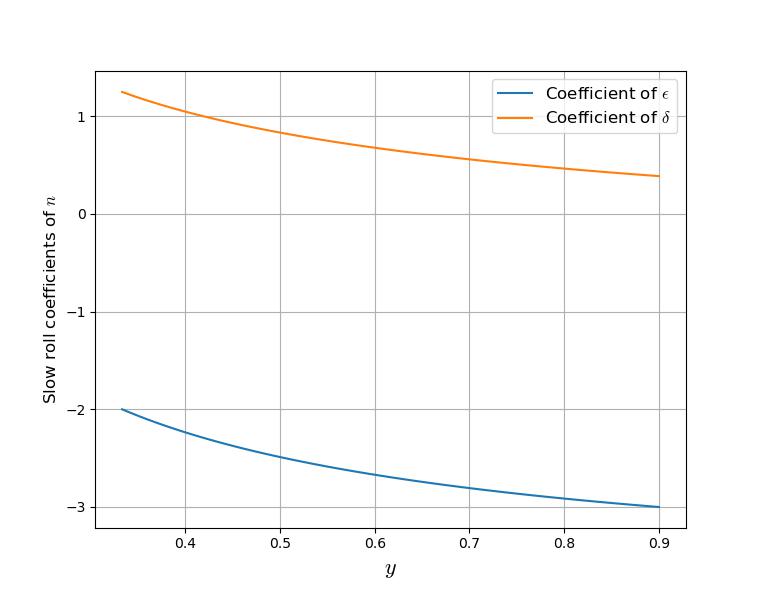}
    \end{subfigure}
    \caption{The figure shows the estimate of the functions involved in the expression for $n$}
    \label{fig:5}
\end{figure*}

We can see from here that the slow roll coefficients are of order $0.1-1$. The smallness of the slow roll parameters will reduce their contribution to the value of $m$ in eq. (\ref{6.45}). The de-Sitter value of $m$ is close to 0 near the boundary value which is $y=1/3$. Choosing a value of $y$ close to this boundary will lead to a very small $m$. In this limit, $h_e$ can be treated as a constant in propertime on super horizon scales. This is the expected behaviour of the gravitational waves after crossing the
horizon, i.e, the oscillation amplitude becomes negligible while comparing with the
wavelength and the wavelength is said to be frozen. If we are to choose a value of $y$ further far from the boundary, the super horizon modes are no longer nearly constant. They will have a growing nature as they go more and more outside the horizon.
The power spectrum is calculated as,
\begin{equation}\label{6.46}
    P(k)=\frac{k^3}{2\pi^2}\sum_e|h_e|^2\propto k^{3-2\nu} \propto k^n
\end{equation}
Thus, the spectral index, $n$, is given as,
\begin{equation}\label{6.47}
    n=3(1-p)+2\epsilon\left(\frac{q-1}{p} \right)+2\delta\left(\frac{Q/3-f}{p} \right)
\end{equation}
The estimate of the functions in the spectral index is shown in fig. (\ref{fig:5}). According to the terminology we mentioned earlier, the de-Sitter value of $n$ corresponds to the part of the spectral index which is independent of slow roll parameters.

For a value of $y$ near the boundary($x=1/3$), the de-Sitter part of $n$ is very small and negative. The slow roll coefficients are of the order 1, but the cumulative effect will be small since they come together with the slow roll parameters. Hence, we can obtain a nearly scale invariant power spectrum in the slow roll limit by choosing $y$ to be close to $1/3$. Note that even for the de-Sitter case, there is only a near scale invariance, unlike in Ref. \cite{Aashish:2020mlw}, where we obtain a perfectly scale invariant power spectrum in the de-Sitter limit. Earlier, we had talked about a constraint which we set in our model, i.e  $m=(\frac{\Dot{\phi}}{\phi}+2H)$. If we are not to set such a constraint and apply the analysis for the $F(H,\phi)$ given in eq. (\ref{6.17}), the tensor modes will have a strong temporal dependence and we get a power spectrum with strong scale dependence in the super horizon limit.
\section{Conclusion}\label{sec:7}
We analyzed the implications of working with an inflation model driven by a triplet of antisymmetric tensor fields. Similar to the previous study on antisymmetric tensor field inflation driven by a single field, slow roll inflation with enough number of e-folds is achieved by incorporating non-minimal couplings with gravity \cite{Aashish:2020mlw}. But the stress-energy tensor we obtained in this case was diagonal. Thus, isotropy is an inherent feature of this model and there is no need to impose additional conditions for ensuring isotropy, unlike the previous studies. The non-minimal coupling strength was constrained by looking at the evolution of the slow roll parameters. Coupling strengths which allowed the slow roll parameters to maintain a small value for at least 70 e folds were selected. We also looked at the energy conditions satisfied by our model. Then we studied the perturbations, initially only in the driving field triplet, $B^{(k)}_{\mu\nu}$, and then in the tensor sector of the metric $g_{\mu\nu}$. The model is free from ghost instabilities. Further, we looked at the primordial gravitational waves that generate from the tensor perturbations in the metric. The velocity of these waves was found to be $c$, which is the velocity of light in vacuum and matches with the recent GW data which constrains the GW speed to around $c$. Although, the GW speed constraints are only valid for astrophysical sources, they are not technically applicable to primordial gravitational waves. But they can be necessitated if we are to let the external fields correspond to dark matter or dark energy. Then we studied the evolution of these tensor modes. Solving, the GW equation, we could see that the modes have an oscillatory behaviour in the subhorizon limit. But, on superhorizon scales these modes were nearly frozen in time. Further, we could get a nearly scale invariant power spectrum.

An obvious next step in this analysis is the study of vector and scalar perturbations, which will be dealt with in the upcoming works. Another possible direction of investigation would be to treat the triplet of fields as dark matter or dark energy and looking at their effects on the recent universe. 

\section{Acknowledgements}
This work is partially supported by DST (Govt. of India) Grant No. SERB/PHY/2021057.
\\\\\\\\\\\\\\\\\\\\\\\
\begin{appendices}
\section{Functions involved in the evolution of slow roll parameters}\label{app:A}
\begin{dmath}\label{4.17}
    f_{1n}=-[\alpha(-2-2y+\alpha+\delta(-1+2y-\epsilon)+\epsilon(2y-2))(-2-6y+(-2+8y)\delta+(1+2y-\epsilon)\epsilon)-(2-6y+\alpha+\delta(1+2y-2\epsilon)+4(-1+y)\epsilon)((-2+8y)\delta^3+\delta^2(-7+(2-12y)\epsilon)+\alpha(-2+2(-1+6y)\delta+(3-6y)\epsilon-2\epsilon^2)-2\epsilon(5-3y+(1-5y)\epsilon+2(-1+y)\epsilon^2)+\delta(2-18y)+3(1+6y)\epsilon+2\epsilon^3-(1+4y)\epsilon^2)]
\end{dmath}
\begin{dmath}\label{4.18}
    f_{1d}=(-2+\epsilon)(\alpha+\delta(-1+4y-\epsilon)+2(-1+y-2y^2+(-1+y)\epsilon))=f_{2d}
\end{dmath}
\begin{dmath}\label{4.19}
    f_{2n}=-2(-1+4y)\delta^3(-2+2y+\epsilon)+\delta^2(14(-1+y)+(11-28y+24y^2)\epsilon+2(-1+6y)\epsilon^2)+2\epsilon(-2(5-8y+3y^2)+(3+9y-10y^2)\epsilon+(5-13y+4y^2)\epsilon^2+2(-1+y)\epsilon^3)+\delta(4-40y+36y^2+(4+48y-36y^2)\epsilon+(-5-24y+8y^2)\epsilon^2+5\epsilon^3-2\epsilon^4)+\alpha(-4(1+3y^2)+8(1-2y+2y^2)\epsilon+(-7 +8y)\epsilon^2+2\epsilon^3-2\delta(2-12y+4y^2+(-1+6y)\epsilon)
\end{dmath}
\end{appendices}
%

%
\bibliographystyle{ieeetr}
\bibliography{refs}
%
%
%

\end{document}